\newtheorem{theorem}{Theorem}

\newtheorem{definition}[theorem]{Definition}

\documentclass{article}

\usepackage{amssymb,amsmath,epsf}
\usepackage{wasysym}

\begin{document}

\title{Loop Variable Representation of Classical Higher Dimensional Gravity and the Hilbert Space Grassmannian}
\author{ Madhavan Venkatesh \\ \\  Centre For Fundamental Research and Creative Education\\ 
Bangalore, India} 
       
\date{May 26, 2013}
\maketitle

\begin{abstract}
In this paper, an attempt is made to represent 5+1 dimensional gravity (via ADM formalism) in 
terms of the loop constructions introduced by the author in a companion paper. The "momenta" and 
"velocity" from the earlier paper, which were proven to be cobordant loops in 6 D; are used as 
the new loop variables. In the process, the Hamiltonian, Diffeomorphism and Gauss constraints 
are written in polynomials of these loop variables. Other constraints such as the "Q" constraint 
and simplicity constraints arise due to greater degrees of freedom. We then undergo the Master 
Constraint treatment to resolve the constraints. Then, a pre-quantum version of the theory is 
examined; and the properties of the Grassmannian of the Hilbert Space are explored.
\end{abstract}

\section{Introduction}
The aim of this paper is to linearize 5+1 dimensional gravity in terms of loop variables. Loops 
are an ideal basis, as, Q, the 'energy' in ordinary dimensions, is the structure of the space in 
6 dimensions. As has already been showed in \cite{mv}, it is possible to use $\tilde{\pi}$ and $
\varpi$ as the ordinary loop variables. In the first part of this paper, a classical treatment 
of 6 D gravity is provided, including the ADM split; and computing the constraints. In the latter 
part, the Calabi-Yau pre-quantum structure due to the loop space $\Omega \mathrm{G}$ is 
examined. It is observed the Calabi-Yau manifold has a one dimensional $\mathit{U(1)}$ K\"{a}hler compactification analogous to Kaluza-Klein theory. A cell decomposition is done to 
retrieve the physics of the $\mathit{U(1)}$, so the exactness of the two-form $\omega$ on each 
of these cells is examined. It follows a priori, that the piece with an exact form, is the manifold of holonomy $\mathit{U(1)}$, with a manifest K\"{a}hler structure.

Note: Some techniques used in this paper are defined in \cite{mv}. Theorems related to some claims made here are also proven. We refer the reader to the above for mathematical preliminaries.

\subsection{Motivation}
Why higher dimensions? The physical motivation of the 'new loop program' is to bring together some aspects of string theory and loop quantum gravity, in a hope to achieve a step towards unification. Analogous to Kaluza-Klein theory, it is intended to recover electromagnetism from the theory at the Calabi-Yau level. Loop groups, prove to be an extremely useful mathematical structure pre and post-quantization. The Calabi-Yau manifold talked of, is manifest as the Grassmannian of the resultant Hilbert Space. From a purely mathematical perspective, it is possible to construct generalized Kac-Moody algebras as central extensions of the loop algebra, by implying the condition that the roots are simple and imaginary. With weaker conditions, it is possible to even construct a Monster Lie algebra. Also of interest is the fact that vector fields on $S^{1}$ correspond to unbounded operators on Hilbert Space. Supersymmetry arises due to the Calabi-Yau structure and due to the fact that the algebra of the vector fields on $S^{1}$ is a monstrous Lie Superalgebra. Another curious usage is the fact that the 'total product' defined in \cite{mv} , providing for the 'energy' in ordinary dimensions, turns out to give the Maurer Cartan structure for higher dimensions (in particular 6). Due to this convenient fact, (see \cite{mv} ) we are enabled to observe gravity in higher dimensions from a simpler perspective and can be sure of the fact that it would reduce to a familiar theory in ordinary dimensions. So, the theory is well motivated by mathematical interest and physical implications.

\subsection{Preliminaries}
We denote by $LG$ the group of real analytic loops. $ \hat{L}G$ is a sub-group of $LG$ whose elements are ordered pairs of cobordant loops, two of which, being the 'momenta' and 'velocity' $\tilde{\pi}$ and $\varpi$ respectively. The group $\hat{L}G$ has an inner product defined by (for two loops $\xi$ and $\zeta$):
$ \xi\circledast\zeta=\left( \xi\oplus\zeta\right) \circ\left( \xi\ominus\zeta\right),$ where $\circ$ is the Chas-Sulliavan product on the cobordism and $\oplus$ and $\ominus$ are defined in terms of it. We denote, by $\eta$, the 'holonomy average', given by $\eta=\frac{1}{2}\left(\oint dx^{a} dx^{b} A_{ab} + \oint dy^{a} dy^{b} A_{ab} \right)$
The Chas-Sullivan product, endowed with a differential operator $\Theta$ and the Moyal-like 'loop bracket' $\left\lbrace\cdot,\cdot\right\rbrace $, renders the homology to $\hat{L}G$; to a Gerstenhaber qua Batalin-Vilkovisky algebra.
We denote by $\Omega\mathfrak{g}$, the space tangent to $\hat{L}G$, that is, the loop algebra, from the universal central extension of a Lie Algebra $\mathfrak{g}$ as $\left[ \mathfrak{g},\mathfrak{g}\right] =\hat{\mathfrak{g}}.$ Let $G$ be the Lie Group and $\mathcal{M}$, its group manifold. We define a flat connection $A$, compatible with the quantization condition:  $A_{ab}=A_{a}A_{b}-A_{b}A_{a}+ A_{\left[ a,b\right] }.$ The structure equation is given by $dA=\int_{\Omega G}\tilde{\pi}\circledast\varpi$

\section{5+1 Split}
We start with a Lorentzian manifold $\mathcal{M}^{6}$ diffeomorphic to  $\Omega G$ $\times$ $
\Sigma$. The metric is decomposed: $q_{ab}=g_{ab} + n_{a}n_{b}.$ We introduce a time-like vector field: $t^{a}=\mathit{N}n^{a}+\mathit{N^{a}}+\acute{
\mathit{N}}n^{a}+\acute{\mathit{N^{a}}},$ 
where $\mathit{N}$ is the lapse function, $\mathit{N^{a}}$ is the shift vector, $\acute{\mathit{N
}}$ is the turn,  $\acute{\mathit{N^{a}}}$ is the twist vector and $n^{a}$ is the unit normal. We define the extrinsic 
curvature $K_{ab}$ by \begin{equation}K_{ab}= -\frac{1}{2}\left\lbrace  \frac{1}{\mathit{N}+ \acute{\mathit{N}}}\left
[ \frac{\partial q_{ab}}{\partial t}- \left( \mathit{N}_{a;b}+ \mathit{N}_{b;a}\right) -\left( 
\acute{\mathit{N}}_{a;b} + \acute{\mathit{N}}_{b;a}\right) \right] \right\rbrace .\end{equation} As in \cite{mv} we have the 5-Ricci scalar $R= \int\tilde{\pi}\oplus\varpi + \int\tilde{\pi}\ominus\varpi.$
The ADM action is given by: 
\begin{equation} S=\int dt\int d^{5}x\sqrt{q}\left( \mathit{N} + \acute{\mathit{N}}\right) \left( R + K^{ab}K_{ab} - K^{2}\right).
 \end{equation}
Computing the quadratic in $K$ as polynomial in the loop variables, we rewrite the action as 
$$ S=\int dt \int d^{5}x \sqrt{q} \left( \mathit{N} + \acute{\mathit{N}} \right)  [  \left( \int\tilde{\pi}\oplus\varpi + \int \tilde{\pi}\ominus\varpi\right)+  \left( \int \tilde{\pi}\circ\varpi\circ\tilde{\pi}\times\int\tilde{\pi}\circ\varpi\circ\varpi\right) $$ \ \begin{equation} - \left(\int\tilde{\pi}\circ\left( \tilde{\pi}\circ\varpi\right) \circ\varpi\right)  ] ,\end{equation}
having substituted for $R$, as well, in terms of the loop variables.
The canonical momenta, conjugate to the connection, is then calculated as:

\begin{equation}
p= \frac{\delta}{\delta A}\left[ \left( \int\tilde{\pi}\oplus\varpi + \int \tilde{\pi}\ominus\varpi\right)+  \left( \int \tilde{\pi}\circ\varpi\circ\tilde{\pi}\times\int\tilde{\pi}\circ\varpi\circ\varpi\right)-  \left(\int\tilde{\pi}\circ\left( \tilde{\pi}\circ\varpi\right) \circ\varpi\right)  \right] ,
\end{equation}
where $\frac{\delta}{\delta A}$ is the Gambini-Pullin connection (contact) functional. 

Correspondingly, the gauge velocity, computed in terms of $K$, and then substituting, just turns out to be ( somewhat analogous to \cite{mv} ) :
\begin{equation}
\vartheta=\int \mathfrak{i}_{n} \left[  \left( \int \tilde{\pi}\circ\varpi\circ\tilde{\pi}\times\int\tilde{\pi}\circ\varpi\circ\varpi\right)-  \left(\int\tilde{\pi}\circ\left( \tilde{\pi}\circ\varpi\right) \circ\varpi\right)  \right],
\end{equation}
where $\mathfrak{i}_{n}$ is the interior product taken with respect to the unit normal $n$.
Importantly, we have, under Poisson brackets,
$$\left\lbrace p^{i}_{a}\left( x\right) , \vartheta^{b}_{k}\left( y\right) \right\rbrace = \delta^{b}_{a}\delta^{i}_{k}\delta^{5}\left( x-y\right) .$$
And, under the 'loop'-Moyal bracket, we have:
$$ \left\lbrace p, \vartheta \right\rbrace_{\mathbb{M}} = p\circ\vartheta + \vartheta\circ p.
$$
\section{Palatini gravity and the new loop Variables}
This section will comprise of casting the above formulation in Palatini form, and representing the derived constraints in terms of the new loop variables.
We start with the action:
\begin{equation}
S=\int dt\int _{\mathcal{M}} d^{6}x  \sqrt{q}  e^{a}_{I} e^{b}_{J} F_{ab}^{IJ} ,
\end{equation}
where $\mathcal{M}$ denotes the space-time manifold of topology $\Omega G \times \Sigma$, $e^{a}_{I}$ are the vielbein and $q_{ab}$ is the space-time metric.
Rewriting the action, (after the space-time split) we get:
\begin{equation}
S=\int dt \int_{\Sigma} d^{6} x \left( \mathit{N}\mathcal{H} - \mathit{N}^{a}\mathcal{H}_{a} + 
\acute{\mathit{N}} \mathcal{Q} - \acute{\mathit{N}}^{a} \mathcal{S}_{a} + p^{a} \mathcal{G}_{a} 
\right),
\end{equation}
where $\mathcal{H}$ is the Hamiltonian constraint, $\mathcal{H}_{a}$ is the Diffeomorphism constraint, $\mathcal{G}_{a}$ is the Gau\ss,constraint, and the $\mathcal{S}_{a}$ are simplicity constraints. We compute the constraints in terms of the loop variables as:

\begin{equation} \mathcal{H}= \int [ \frac{\delta}{\delta A}\left[ \left( \int\tilde{\pi}\oplus\varpi + \int \tilde{\pi}\ominus\varpi\right)+  \left( \int \tilde{\pi}\circ\varpi\circ\tilde{\pi}\times\int\tilde{\pi}\circ\varpi\circ\varpi\right)-  \left(\int\tilde{\pi}\circ\left( \tilde{\pi}\circ\varpi\right) \circ\varpi\right)  \right] ]  \end{equation} \ $$ \oplus [ \int \mathfrak{i}_{n} \left[  \left( \int \tilde{\pi}\circ\varpi\circ\tilde{\pi}\times\int\tilde{\pi}\circ\varpi\circ\varpi\right)-  \left(\int\tilde{\pi}\circ\left( \tilde{\pi}\circ\varpi\right) \circ\varpi\right)  \right] ].                           $$      

\

\begin{equation} \mathcal{H}_{a}= \frac{\delta}{\delta A^{a}}\int [ \frac{\delta}{\delta A}\left[ \left( \int\tilde{\pi}\oplus\varpi + \int \tilde{\pi}\ominus\varpi\right)+  \left( \int \tilde{\pi}\circ\varpi\circ\tilde{\pi}\times\int\tilde{\pi}\circ\varpi\circ\varpi\right)-  \left(\int\tilde{\pi}\circ\left( \tilde{\pi}\circ\varpi\right) \circ\varpi\right)  \right] ]  \end{equation} \ $$ \ominus [ \int \mathfrak{i}_{n} \left[  \left( \int \tilde{\pi}\circ\varpi\circ\tilde{\pi}\times\int\tilde{\pi}\circ\varpi\circ\varpi\right)-  \left(\int\tilde{\pi}\circ\left( \tilde{\pi}\circ\varpi\right) \circ\varpi\right)  \right] ].                           $$

\

\begin{equation}
\mathcal{G}_{a} = D_{a}[ \frac{\delta}{\delta A}\left[ \left( \int\tilde{\pi}\oplus\varpi + \int \tilde{\pi}\ominus\varpi\right)+  \left( \int \tilde{\pi}\circ\varpi\circ\tilde{\pi}\times\int\tilde{\pi}\circ\varpi\circ\varpi\right)-  \left(\int\tilde{\pi}\circ\left( \tilde{\pi}\circ\varpi\right) \circ\varpi\right)  \right] ] 
\end{equation}

\

\begin{equation} \mathcal{Q}= \int [ \frac{\delta}{\delta A}\left[ \left( \int\tilde{\pi}\oplus\varpi + \int \tilde{\pi}\ominus\varpi\right)+  \left( \int \tilde{\pi}\circ\varpi\circ\tilde{\pi}\times\int\tilde{\pi}\circ\varpi\circ\varpi\right)-  \left(\int\tilde{\pi}\circ\left( \tilde{\pi}\circ\varpi\right) \circ\varpi\right)  \right] ]  \end{equation} \ $$ \circledast [ \int \mathfrak{i}_{n} \left[  \left( \int \tilde{\pi}\circ\varpi\circ\tilde{\pi}\times\int\tilde{\pi}\circ\varpi\circ\varpi\right)-  \left(\int\tilde{\pi}\circ\left( \tilde{\pi}\circ\varpi\right) \circ\varpi\right)  \right] ].                           $$

\pagebreak
\begin{equation} \mathcal{Q}= \int [ \frac{\delta}{\delta A}\left[ \left( \int\tilde{\pi}\oplus\varpi + \int \tilde{\pi}\ominus\varpi\right)+  \left( \int \tilde{\pi}\circ\varpi\circ\tilde{\pi}\times\int\tilde{\pi}\circ\varpi\circ\varpi\right)-  \left(\int\tilde{\pi}\circ\left( \tilde{\pi}\circ\varpi\right) \circ\varpi\right)  \right] ]  \end{equation} \ $$ \oplus [ \int \mathfrak{i}_{n} \left[  \left( \int \tilde{\pi}\circ\varpi\circ\tilde{\pi}\times\int\tilde{\pi}\circ\varpi\circ\varpi\right)-  \left(\int\tilde{\pi}\circ\left( \tilde{\pi}\circ\varpi\right) \circ\varpi\right)  \right] ] $$ \ $$  [\circ \frac{\delta}{\delta A^{a}}\int [ \frac{\delta}{\delta A}\left[ \left( \int\tilde{\pi}\oplus\varpi + \int \tilde{\pi}\ominus\varpi\right)+  \left( \int \tilde{\pi}\circ\varpi\circ\tilde{\pi}\times\int\tilde{\pi}\circ\varpi\circ\varpi\right)-  \left(\int\tilde{\pi}\circ\left( \tilde{\pi}\circ\varpi\right) \circ\varpi\right)  \right] ]  $$ \ $$ \ominus [ \int \mathfrak{i}_{n} \left[  \left( \int \tilde{\pi}\circ\varpi\circ\tilde{\pi}\times\int\tilde{\pi}\circ\varpi\circ\varpi\right)-  \left(\int\tilde{\pi}\circ\left( \tilde{\pi}\circ\varpi\right) \circ\varpi\right)  \right] ]] $$

\

\begin{equation} \mathcal{S}= \int [ \frac{\delta}{\delta A}\left[ \left( \int\tilde{\pi}\oplus\varpi + \int \tilde{\pi}\ominus\varpi\right)+  \left( \int \tilde{\pi}\circ\varpi\circ\tilde{\pi}\times\int\tilde{\pi}\circ\varpi\circ\varpi\right)-  \left(\int\tilde{\pi}\circ\left( \tilde{\pi}\circ\varpi\right) \circ\varpi\right)  \right] ]  \end{equation} \ $$ \circ [ \int \mathfrak{i}_{n} \left[  \left( \int \tilde{\pi}\circ\varpi\circ\tilde{\pi}\times\int\tilde{\pi}\circ\varpi\circ\varpi\right)-  \left(\int\tilde{\pi}\circ\left( \tilde{\pi}\circ\varpi\right) \circ\varpi\right)  \right] ].                           $$

It is apparent to see that the constraint $\mathcal{Q}$ is polynomial in the other constraints, and particularly is a composition of the Hamiltonian and Diffeomorphism constraints. It is also interesting that most of the constraints contain the same initial expression, but the products between them, and the differential operators acting on them provide for the difference. The constraints have been expressed in a terse, foreign fashion, but they can be resolved by the introduction of the Master Constraint.
We write the master constraint for the theory as follows:
\begin{equation}
\mathbf{M}= \int_{\Sigma} d^{5}x \frac{\left[ \mathcal{Q}^{2}\right] }{\sqrt{q}}.
\end{equation}

\paragraph{Discussion}
It is known that the Gambini-Pullin derivative is a a generator of diffeomorphisms, and the loop derivative, a generator of the group of loops. Now, the Master Constraint is quadratic in the $\mathcal{Q}$ constraint instead of the traditional Hamiltonian constraint. Now, the Master constraint will be polynomial in $p$ and $\vartheta$ via the 'total product', which provides for the Maurer-Cartan structure of General Relativity. The theory remains to be diffeomorphism invariant and locally Lorentz invariant, as with LQG, but, here, the construct of the dynamical variables eases the ability to 'retrieve' GR at the classical limit of the quantum theory.
The loop space, being dealt with, can be quantized by methods such that a definite classical limit can be obtained. Berezin-Toeplitz approach is quite suitable for this primer. Some further ambiguities in LQG in the context of this theory, eg. the Hamiltonian Constraint will be discussed along with quantization,in following papers.

\pagebreak

\section{Topological aspects}
The Loop Space $\Omega G$ in question, has a symplectic and manifest K\"{a}hler structure. Adding to that, the fact that we are dealing with a Ricci-flat theory, it is bound to yield a Ricci-flat K\"{a}hler manifold (ie. Calabi-Yau), post-quantization, as the Hilbert Space Grassmannian.

\paragraph{Quantizability }
Our loop space and the dynamics on it have been constructed pertaining to certain quantization conditions, ie. the connection, and the possibility of constructing a holomorphic line bundle. We simply require the K\"{a}hler form to be the curvature of the holomorphic line bundle on $\Omega G$, enabled with projection to the Hilbert Space. Henceforth, we shall consider a pre-quantum theory with the Hilbert Space $H=L^{2}\left( S^{1},\mathbb{C}\right) $, with polarization $H=H_{+}\oplus H_{-}.$  We define the Grassmannian of the Hilbert Space along te lines of \cite{ps}

\begin{definition}
The Grassmannian of the Hilbert Space, $Gr(H)$ is 
the set of all closed subspaces $W$ of $H$, such that: \\

i. the projection $p_{+} : W \rightarrow  H_{+}$ is a Fredholm operator, \\

ii.the projection $p_{-} : W \rightarrow H_{-} $ is a Hilbert-Schmidt operator.
\end{definition}

Now, we investigate the K\"{ahler} (Calabi-Yau) structure of the Loop space and the would-be Grassmannian.

On $\Omega G$, we have a K\"{a}hler metric, with a volume form given by :
$$ \omega^{6}= \frac{i^{6}}{6 !}det\left( g_{ab}\right) dz^{1} \wedge dz^{2}  ... \wedge dz^{6}.$$

We have, the Ricci curvature as before (in terms of loops) 
$$R= \int \tilde{\pi}\oplus\varpi + \int \tilde{\pi} \ominus\varpi .$$
We write down, the average of the scalar curvatures (total) :
$$\breve{R}= \int\tilde{\pi}\circledast\varpi .$$
Now, we are ready to define the Calabi energy:
$$ \mathtt{C} = \int_{\Omega G}\left( R - \breve{R}\right) ^{2} \omega .$$

At the potential level, we have, the Calabi flow, of the form:
$$\frac{\partial \varphi}{\partial t}= R - \breve{R}.$$

Now, we have an explicit energy-flow function.

$$\mathcal{E}\left( \tilde{\pi},\varpi\right) =
\mathtt{C}_{\left( \tilde{\pi},\varpi\right) }.$$
We can also write an operator $i \frac{d}{d \theta}$, being the expected value of the quantum Calabi-energy operator ( an unbounded Hermitian operator) on Hilbert Space.

\subsection{U(1) holonomy}

Post-quantization, the Grassmannian turns out to be a Calabi-Yau 6-fold with a $U(1)$ compactification, analogous to Kaluza-Klein theory. The compactified piece is K\"{a}hler with a $U(1)$ holonomy. in order to recover the physics of the $U(1)$, it is a necessity that we must perform surgery. There are a few ways to go about this, including the Sch\"{u}bert cell decomposition, CW complexes, handle-body decomposition and the Birkhoff and Bruhat decompositions. We undego a Heegard-like splitting for this six-fold, intending to split it into 5 five Calabi-Yau pieces and one  K\"{a}hler piece (with $U(1)$ holonomy).
We define a Morse function, called the 'stitch function' $ f: Gr(H) \rightarrow Gr(H)$ :
$$f:Gr(H) \rightarrow Gr(H) = \mathcal{M}_{1}\subset\mathcal{M}_{2}\subset\mathcal{M}_{3}\subset\mathcal{M}_{4}\subset\mathcal{M}_{5}\subset\mathcal{M}_{6}.
$$
The K\"{a}hler form on $Gr(H)$, gets split as follows :
$$\Omega=\Omega_{1}\wedge\Omega_{2}\wedge\Omega_{3}\wedge\Omega_{4}\wedge\Omega_{5}\wedge\Omega_{6}.
$$
We can now check for the closure of the K\"{a}hler form on each of these pieces, by acting the exterior differential operator on $\Omega$. Consider,  the only form that doesn't close under the exterior derivative to be $\mathcal{M}_{6}$. Now, we can separate these pieces , by introducing another Morse function (which we shall call the 'smear function'): $\check{f}$. The generically defined function, 'separates out' the pieces, by changing the spaces to space modulo the density on it. (It can be assumed that integration on each stratum is variant), ie.
$$ \check{f} : \mathcal{M}_{n} : \mathcal{M}_{n} \rightarrow \mathcal{M}_{n} = \frac{\mathcal{M}_{1}}{\mu_{1}} + \frac{\mathcal{M}_{2}}{\mu_{2}} + .....
\frac{\mathcal{M}_{6}}{\mu_{6}}.$$

Now, since the pieces have been separated out, we can cell-stratify by defining another Morse function (called the 'sew function') $\breve{f}$. This function stratifies each cell, to give back the Calabi-Yau structure that was present before the operations of the Morse functions.
$$\breve{f}: \mathcal{M}_{n} \rightarrow Gr(H) = \mathcal{M}_{1}\cup\mathcal{M}_{2}\cup  ... \mathcal{M}_{6}=Gr(H).$$

It is apparent that the new Calabi-Yau six-fold is not exactly the same as the one we had earlier, but, is in fact, dual to it (due to the stratification). Thus, by virtue of this, we have a twisted form of mirror symmetry.
\pagebreak
\section{Conclusion}
In this paper, we have performed a 5+1 split of gravity and calculated the constraints for it, in terms of the new loop variables. We introduced a new, more 'fundamental' constraint $\mathcal{Q}$, as a composition of the Hamiltonian and Diffeomorphism constraints. As we engage in dimensions above the ordinary, we have had to deal with Simplicity constraints as well. This formalism attempts to reconcile some aspects of string theory and loop quantum gravity to do away with some ambiguities in the latter. In the second part of this paper, we performed surgery on the Grassmannian of the prospective Hilbert Space, to manifest the symmetry of the $U(1)$. Ambiguities in LQG from the context of this theory and quantization of this theory will be the subject of forthcoming papers.

\section{Acknowledgments}

This work was carried out at the Center For Fundamental Research And Creative
Education (CFRCE), Bangalore, India, under the guidance of Dr B S Ramachandra
whom I wish to acknowledge.  I would like to acknowledge the Director Ms. Pratiti B R for creating the highly charged research atmosphere at CFRCE. I would also like to thank my fellow researchers Manogna H Shastry,Vasudev Shyam, Karthik T Vasu and Arvind Dudi.


\begin{thebibliography}{8}

\bibitem{mv} Venkatesh, Madhavan.
\textit{An Algebraic Topological Construct of Classical Loop Gravity and the prospect of Higher Dimensions}. arXiv:1305.0383

\bibitem{ps} Pressley, Andrew , Segal, Graeme. \textit{Loop Groups}. Oxford Mathematical Monographs. 1986.
\bibitem{GP} Gambini, Rodolfo , Pullin, Jorge.
\textit{Loops, Knots, Gauge theories and Quantum Gravity}. Cambridge University Press. 2000.


\bibitem{tt1} Bodendorfer, Norbert , Thiemann, Thomas, Thurn Andreas
\textit{New Variables for Classical and Quantum Gravity in all Dimensions I. Hamiltonian Analysis} arXiv: 1105.3703

\bibitem{tt1} Bodendorfer, Norbert , Thiemann, Thomas, Thurn Andreas
\textit{New Variables for Classical and Quantum Gravity in all Dimensions I. Lagrangian Analysis} arXiv: 1105.3704

\bibitem{ch} Chen, Xiuxiong , He Weiyong
\textit{On the Calabi flow} arXiv: math/0603523


  



\end{thebibliography}
\end{document}